\documentclass[aps,12pt,floatfix,tightenlines,showkeys,showpacs]{revtex4}
\usepackage{epsfig}
\usepackage{graphicx}
\usepackage{amssymb}
\newcommand{\bfr}{{\bf r}}

\newcommand{\ben}{\begin{displaymath}}
\newcommand{\een}{\end{displaymath}}
\newcommand{\be}{\begin{equation}}
\newcommand{\ee}{\end{equation}}
\newcommand{\bea}{\begin{eqnarray}}
\newcommand{\eea}{\end{eqnarray}}

\newcommand{\eq}[1]{Eq.~(\ref{#1})}

\newcommand{\bfp}{{\bf p}}

\begin{document}

\title{\bf  \hskip10cm NT@UW-10-06\\{Color Transparency at COMPASS energies}}

\author{ Gerald A.  Miller$^1$, Mark Strikman$^2$}

\affiliation{
$^1$ University of Washington, Seattle, WA 98195-1560\\
$^2$ Pennsylvania State University, University Park, PA 16802}

\date{\today}

\begin{abstract}

{Pionic quasielastic knockout of protons from nuclei at 200 GeV show very large effects of color transparency as  $-t$ increases from 0 to several GeV$^2$.  Similar effects are  expected for quasielastic photoproduction of vector mesons.}
\end{abstract}

\maketitle     
 
\section{Introduction}

In the very special situation of high-momentum-transfer coherent processes 
the strong interactions between hadrons and nuclei 
can be extinguished, causing  shadowing to disappear and the
nucleus to become quantum-mechanically transparent. This phenomenon is known as
color transparency \cite{ct,Frankfurt:1994hf,Miller:2007zz,fs}. In more technical language, color transparency is the
vanishing of initial and final-state interactions, predicted by QCD to occur in
high-momentum-transfer quasi-elastic nuclear reactions. 
In these reactions, the scattering amplitudes consist of a sum of terms involving different
intermediate states and the same final state. Thus 
 one adds different contributions to obtain the scattering
amplitude. Under such conditions the effects of gluons emitted by small
color-singlet systems tend to cancel \cite{lonus} and could nearly vanish.  Thus color transparency is also known as
color coherence. 

The important dynamical question is whether or not 
small color-singlet systems, often referred to as
point-like configurations (PLC's),
are  produced as intermediate states in high momentum transfer reactions.
Perturbative QCD predicts that a PLC
is formed 
in many two-body hadronic processes
at very large momentum transfer \cite{ct,liste}. However, PLC's can also 
be formed under
non-perturbative dynamics \cite{fms1,fms2}. Therefore 
measurements of color transparency are important for clarifying the dynamics
of bound states in QCD.

Observing
color transparency requires that a PLC is formed and that the energies are
high enough so that  the PLC does not expand completely 
to the  size of a physical hadron  while
traversing the target \cite{ffs,jm,bm}. The frozen approximation must be valid.

A direct observation of high-energy color transparency in the
$ A $-dependence of diffractive di-jet production by pions was reported in \cite{dannyref}.
The results were in accord with the prediction of \cite{Frankfurt:1993it}. 
See also \cite{Frankfurt:1999tq}.
Evidence for color transparency 
(small hadronic cross-sections) has been observed in other 
types of processes, also occurring at high energy:
in the A-dependence of
$J/\psi$ photoproduction \cite{e691}, 
in the $ Q^2 $-dependence of the $ t $-slope of diffractive $ \rho^0 $
production in deep inelastic muon scattering
(where $ Q^2 $ is the invariant mass of the virtual photon and
$ t $ denotes the negative square of the 
momentum transfer from the virtual photon
to the target proton),
 and in
the energy and flavor dependences of vector
meson production in $ ep $ scattering at HERA \cite{ha}.
For all of these processes the energy is high enough so that 
the produced small-size configuration does not expand significantly as it makes its way out 
of the nucleus.

For 
  hard, high-energy processes 
in which  a small dipole is produced (pion diffraction into two jets) 
or the initial state is highly localized (exclusive production of mesons for large values of 
$Q^2$), one can prove
  factorization theorems  which allow the scattering amplitude to be represented 
as the product of the generalized parton densities of the target, hard interaction block, and wave functions 
of projectile and the  final system in the frame where they have high momenta 
\cite{Frankfurt:1993it,Frankfurt:2000jm,Brodsky:1994kf,Collins:1996fb}. The proofs require 
  the  color transparency property of  perturbative QCD,
 understood in the sense of the suppression, $\propto d^2$, of  
multiple interactions of a color electric dipole moment.
Note that  the  definition of  color transparency does not simply correspond
to the nuclear amplitude being $A$ times the nucleonic amplitude because
both the gluon, $G_A$, and quark sea $S_A$, densities   may depend upon the nuclear environment. Instead, color transparency
corresponds to  the dominance of the leading twist term in the relevant scattering amplitude
 \cite{Frankfurt:1993it}.

At the energies available at JLAB and BNL expansion effects do occur.  
Experimental studies of high momentum transfer 
 processes in $ (e,e'p)$ and (p,pp) reactions
 have so far failed to produce
convincing evidence of color transparency\cite{eva,Aclander:2004zm,slac,cebaf}.
 First data on the  reaction $A(p,2p)$ at large scattering angles
 were obtained at BNL. 
They were followed  by the dedicated experiment EVA. The final results of 
EVA \cite{Aclander:2004zm}
 can be summarized as follows.
An eikonal approximation calculation  agrees with  data for  
$\mbox{p}_p$=5.9 GeV/c, and  the transparency increases significantly for momenta up to about 
$\mbox{p}_p$= 9 GeV/c. Thus it seems that 
 momenta of the incoming proton $\sim $
10 GeV are  sufficient to  significantly suppress  expansion effects. 
Therefore one can use proton projectiles 
with  energies above $\sim $10 GeV to study other aspects of the strong interaction dynamics.
But the observed drop in transparency  for values of $\mbox{p}_p$ ranging from 
11.5 to  14.2 GeV/c 
represents a problem for all current models, including  
\cite{Brodsky:1987xw,Ralston:1990jj,Jennings:1990ma,Jennings:1992hs,Frankfurt:1994nn}
 because of its broad range in energy.
 This suggests that  leading-power  quark-exchange mechanism for
 elastic scattering dominates only at very large energies.

It is natural to expect that it is easier to observe color transparency  for
 the interaction/production  of mesons than 
for baryons because   only two quarks have to come close together. 
A high resolution pion production experiment
 reported 
evidence for the onset of CT \, \cite{:2007gqa} at Jefferson Laboratory
in the process $eA\to e\pi^+ A^*$. 
 The experimental results  agree well with predictions of \cite{Larson:2006ge} and
 \cite{Cosyn:2007er} which predict small, but significant effects of color transparency.

In the present note we observe that studying the quasielastic knockout of a proton from a nucleus by the  high energy pions available at COMPASS 
offers a unique opportunity to observe the pionic PLC and even to study the 
its cross section as a function of $-t$.  Our analysis applies also to another reaction which can be studied by COMPASS - quasielastic production of vector mesons in muon - nucleus  interactions. 
 The theory is presented in Sect.~II, and the results in Sect.~III. Kinematic considerations, which show that the proton emission angle is large enough for proton detection, are presented in Sect.IV.

\section{Theory for the Nuclear $\pi,\pi p$ Reaction at High Momentum Transfer}
\label{sec:gen}

It is worthwhile to discuss color transparency for  quasi-elastic scattering or pions from an initially bound proton.
The basic postulate is that at large center-of-mass angles, where $-t>-t_0\sim 1 {\rm GeV}^2$ the reaction proceeds
by components PLC of the pion wave function in which the quarks are closely separated.
At high energies, where the space-time evolution of small-sized PLC 
wave packets is slow, one can introduce a notion of the cross section of scattering of a small dipole configuration (say $q\bar q$) of transverse size $d$ on the nucleon \, \cite{Frankfurt:1993it, Blaettel:1993rd} which in the leading log approximation is given by \, \cite{Frankfurt:2000jm}
\begin{equation}
\sigma(d,x_N)= {\pi^2\over 3} \alpha_s(Q^2_{eff}) d^2\left[xG_N(x,Q^2_{eff}) +2/3 xS_N(x, Q^2_{eff})\right],
\label{pdip}
\end{equation}
where $Q^2_{eff} = \lambda/d^2, \lambda= 4 \div 10\;, x=Q^2_{eff}/s$, with
$s$ the invariant energy of the dipole-nucleon system,
 and  $S$ is the  sea quark distribution  for quarks making up the dipole. Matching description of $\sigma_L$ in momentum and coordinate space leads to $\lambda\sim 9$. However sensitivity to the value of $\lambda$ for small $d$ is small.
 At the same time use of a smaller value of $\lambda \sim 4$ allows to make a smooth extrapolation to $\sigma(d,x_N)$ for large dipole sizes. 
 The  difference between \eq{pdip}  and the simplest two gluon exchange model \cite{Gunion:1976iy} is significant for large values of $x$ for which $x$ is very small. An alternative earlier estimate is based on perturbative QCD  and which assumes a smooth matching with the soft regime yields
\cite{Farrar:1988me}
\bea\sigma(d,x_N)\approx\sigma_{PLC}\equiv \sigma_{\rm tot}(p)\frac{n^2\langle  k_t^2 \rangle}{Q^2_{eff}},\;d^2\sim {1\over Q^2_{eff}},\label{plcp}\eea 
as the cross section for the
initially-produced PLC, of momentum $p$, with 
 $n = 2$ for the pion, $n=3$ for the proton 
and $\langle  k_t^2 \rangle^{1/2} \simeq$0.35 GeV.   

The advantage of COMPASS is that if 
 PLCs are involved in a large-$|-t|$, high-energy process, 
 such configurations  move
 through the nucleus without changing their size. Thus there is an opportunity to test the approximations \eq{pdip} and \eq{plcp}.

Next apply these ideas to the process $\pi p\rightarrow \pi p$ on protons initially bound in a nucleus. For a cm scattering angle $\theta_c$ the invariant momentum transfer $t$ is given by
\begin{eqnarray}
-t=4p_c^2\sin^2(\theta_c/2)=Q^2_{eff},
\end{eqnarray}
At COMPASS $p_c^2\approx 100 \;{\rm GeV}^2$ so $-t$ changes from 0 to 10 GeV$^2$ as $\theta_c$ changes from 0 to about 0.35.  It is also  important to observe that $-t$ plays the role of $Q^2_{eff}$ that appears in \eq{pdip} and \eq{plcp}. 
The kinetic energy  of the outgoing proton varies from 0 to  about 5 GeV over that same range. The momentum of the proton must be at least 1 GeV/c for our considerations to be relevant, so we focus on $-t$ greater than about 1 GeV$^2$.

At Jefferson Lab energies the PLC expands while it moves through the nucleus.  This complication is avoided at COMPASS. The pionic PLC easily transverses the nucleus without expanding. The proton may or may not be initially produced as a PLC. If it is produced as a PLC it will expand as it moves through the nucleus. In the advent of expansion
$\sigma_{PLC} $ of \eq{plcp}  is replaced by an effective cross section,
$\sigma_{eff}$, which  takes the changing size of the wave packet  into account.
  The effective interaction
 contains two parts, one for a propagation distance $l$  less than a length $l_h$ 
describing the interaction 
of the expanding PLC, 
another, for larger values of $l>l_h$
describing the final state
 interaction of the physical particle.  We use the expression \cite{Farrar:1988me}
\bea \sigma_{\rm eff}
(p,l) = \sigma_{\rm tot}(p)
 \left[\left(\frac{n^2\langle  k_t^2 \rangle}{Q^2} +
 \frac{l}{l_h}(1 - \frac{n^2 \langle  k_t^2 \rangle}{Q^2}) \right)\theta(l_h-l) +\theta(l-l_h)\right],
\label{sigplc}  \eea
where $l=|\bfp\cdot{\bf l}/p|$ where $\bfp$ is the momentum and ${\bf l}$ is the displacement from the point where the hard scattering occurs.  The quantity $l_h = 2 p/ \Delta M^2$, with $\Delta M^2=0.7{\rm GeV}^2$ for pions.
The prediction that the interaction of the PLC will be approximately proportional to
the propagation distance $l$ for $l<l_h$ is called the quantum diffusion model. The length $l_h$ controls the physics. The conventional approach of Glauber theory is achieved as $l_h$ approaches 0.
For pions of momentum 200 GeV/c $l_h$ is much larger than the diameter of any stable nuclear target.
For protons, the value of $\Delta M^2$ could be higher than that for pions, and the momentum is typically 
 2-3 GeV/c, depending on the value of $-t$.   Here we take $\Delta M^2 $ to be the same for pions and protons.  The large effects of color transparency that we will observe are mainly due to pionic PLCs, so the value of $\Delta M^2 $ for protons is not very important.

The transparency $T_A$ is  defined here as the ratio of the observed nuclear $\pi,\pi p $ cross section
to $A$ (the nucleon number) times the cross section on a free nucleon ${d\sigma\over dt}$, with
perfect transparency occurring for $T_A\rightarrow1$:
\bea T_A(\bfp_0,\bfp_1,\bfp_2)&\equiv&\frac{{d\sigma_A\over dt}}{A {d\sigma\over dt}}.
\eea
The nuclear  transparency $T_A$  is given by
\bea T_A(\bfp_0,\bfp_1,\bfp_2)\approxeq&\int d^3 \rho_A(r){\cal P}_0(\bfp_0,\bfr){\cal P}_1(\bfp_1,\bfr){\cal P}_2(\bfp_2,\bfr)\label{ta}.
\eea
The survival probability ${\cal P}_i(\bfp_i,\bfr) $ for a hadron of momentum $\bfp_i$ is given by
\begin{eqnarray}
{\cal P}_i(\bfp_i,\bfr))=\exp[-\int_{\rm path} dl \;\sigma_{eff}(\bfp_i,l)].
\end{eqnarray}

In the absence of the effects of color transparency,  one expects that 
Glauber theory would provide a reasonable description of the data. In this case $l_c$ is set to 0 for both pions and protons.  We take the  nuclear density to 
\bea \rho(r)={\rho_0\over 1+e^{r-R\over a}},\eea
with $R=1.1 A^{1/3}$ fm, and $a$=0.54 fm, with $\rho_)$ chosen to normalize the density to the nucleon number, $A$.
\section{Results}

\begin{figure}
\epsfig{file=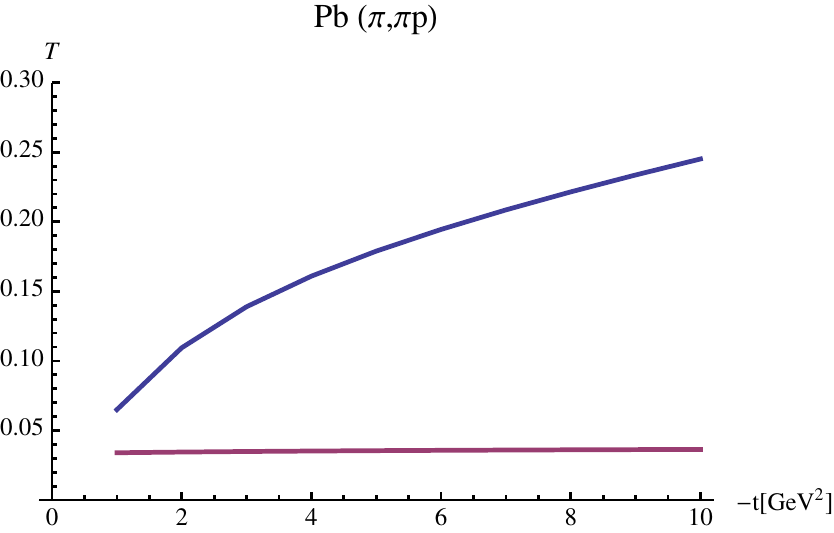, width=12.0cm}
\caption{(Color online) 
The transparency for the $\pi,\pi p$ reaction on $^{208}$Pb The   blue curve includes the effects of color transparency. The lower purple curve represents the effects of the Glauber calculation.
 }% 
\label{Trans}\end{figure}

\begin{figure}
\epsfig{file=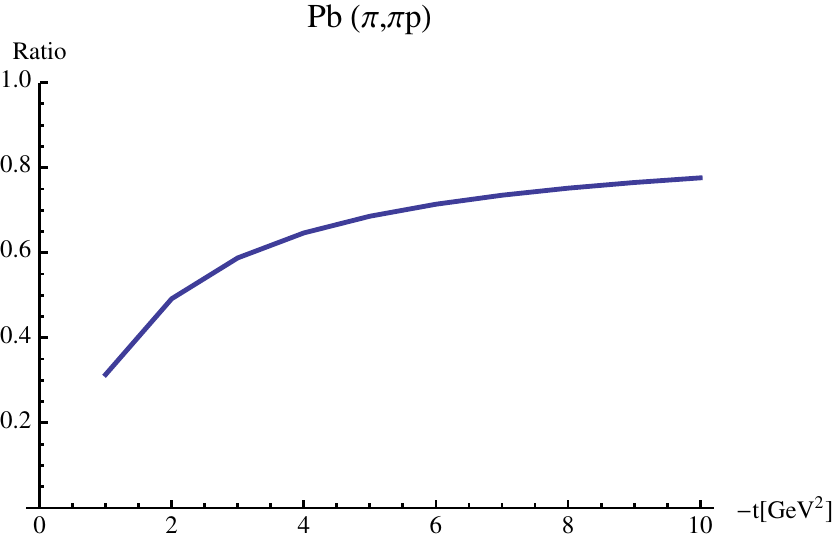, width=12.0cm} 
\caption{(Color online) 
Ratios of  transparency full to plane wave pion for the $\pi,\pi p$ reaction on $^{208}$Pb The upper  blue curve includes the effects of color transparency.}
\label{transratio}\end{figure}
\begin{figure}
\epsfig{file=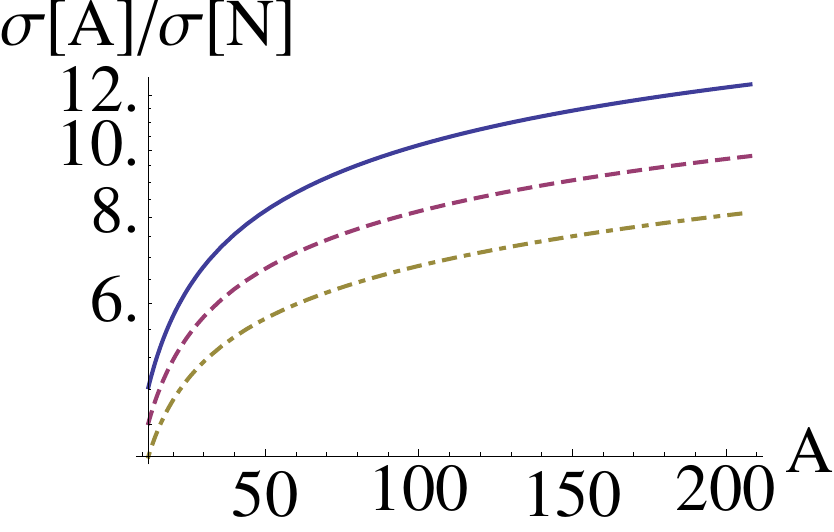,width=12.0cm} 
\caption{(Color online) 
Ratios of  nuclear $\sigma(A)$ to $\sigma(N)$ for three different values of the effective cross section:25 mb (dot-dashed),20 mb (dashed), 15 mb (solid).
 }% 
\label{sigmasigeff}\end{figure}

Figure~\ref{Trans} shows the transparency of \eq{ta}  for $^{208}$Pb with the effects of color transparency and in the Glauber calculation
($l_h=0)$. 
There is a gigantic effect predicted by our  formula \eq{sigplc}. For pions the effective cross section is given by \eq{plcp}
and varying $-t$ has a big effect on the survival probabilities. The proton is strongly influenced by the final state interactions. 

This  effect of the proton final state interactions is  illustrated in Fig.\ref{transratio}. This figure displays  the ratio of $T_A$ of \eq{ta} to the same quantity computed by setting the pion $\sigma_{eff}$ to zero. The large ratio seen indicates 
 a large range of values of $-t$ for which the nucleus is nearly completely transparent to pions.
 
The above results are driven by \eq{sigplc}. However, there is no independent information about this quantity as it enters pion-nucleon elastic scattering.
Hence we explore the sensitivity of the transparency as a function of A to the variation of the the strength of the interaction of the pion with the nucleon.  This is shown in Fig.~\ref{sigmasigeff}.  If $\sigma_{eff}$ is reduced from the full value of 25 mb to 15 mb, one can observe a strong change of the transparency. Hence one would be able to observe even a relatively modest squeezing of the pion wave function well before the full color transparency is reached.

Our results are also applicable  to the process of large -t photoproduction of vector mesons from  nuclei, like 
$\gamma +A \to \rho^0 +N + (A-1)^*$. Indeed, for small -t  it was established a long time ago that the vector dominance model describes well the $\rho$-meson photoproduction with $\sigma_{\rho N}= \sigma_{\pi N}$.  Squeezing in this case should be similar or even stronger than in the pion case due to a singular behavior of the photon wave function at small transverse separations. 
 Note here that the recent studies have suggested that  in the regime 
 when the momentum transfer is larger than the hardness scale of the reaction, the elastic cross section should be energy independent in a wide energy range \cite{Blok}. Inspecting the recent data on the $\gamma +p \to \rho^0 +p$ reaction \cite{List:2009pb}  we notice that  the data are consistent with cross section being energy independent starting with  $-t \ge  0.7 \div 0.8 \mbox{GeV}^2$. 

\section{Kinematic considerations}

Let us analyze the pattern of the emission of the protons which determines requirements on the recoil detector.
The specific feature of  high energy kinematics is that the ``minus" component of the momentum of the struck proton is conserved as the ``minus" components of the initial and final pion are very small - the difference is of the order $-t/s$. 
Hence  the four momentum of the final proton satisfies the condition
\bea
\alpha= (\sqrt{m_N^2+\vec{p}^2} - p_3 )/m_N,\; p_t= q_t +k_t,
\eea
where  $-t=q_t^2$, the light cone fraction $\alpha $ is typically  within the range 
 $|\alpha - 1|\le 0.2$
and 
$k_t$ is the transverse momentum of the struck nucleon in the initial state (typically $\le 0.2 \;\mbox{GeV/c}$). 
\bea
p_3  =   m_N/2 (\alpha^{-1} -\alpha) +  {p_t^2  \over  2\alpha m_N}.
\eea
The first term is the right hand side is  small  as compared to the second term which is 
of the order $-t/2m_N$.
Hence the emission angle (relative to the beam direction) is approximately given by 
\bea
\theta_{em}= \tan^{-1} ( p_3/p_t) \approx \tan^{-1} (p_t/2\alpha m_N).
\label{thet}
\eea
One can see from Eq.\ref{thet} that Fermi motion leads to a modest smearing of the emission angle in the $t$ range we discuss. Also the angle $\theta_{em}$    remains large in the whole range we discuss, which simplifies detection of such protons.\begin{figure}
\epsfig{file=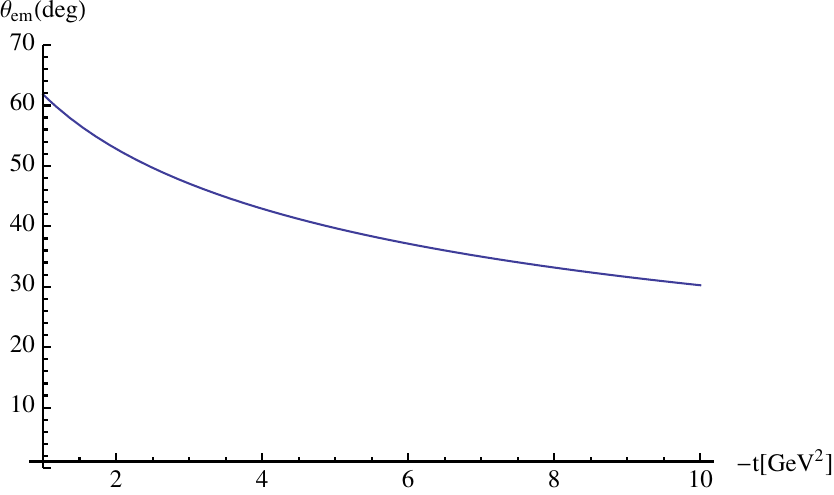,width=12.0cm} 
\caption{(Color online) Proton emission angle $\theta_{em}$ as a function of $-t$.
Exact kinematics are used for a proton initially at rest.
 }% 
\label{angle}\end{figure}\section{Summary}
\label{sec:sum}
We conclude that a measurement of the transparency in the pion quasielastic scattering off nuclei in the COMPASS kinematics may allow to observe a novel color transparency phenomenon. Parallel studies using  quasi real photon production in $\mu +p $ scattering which will be feasible with COMPASS also look promising.

\section*{Acknowledgments}
 This research was supported by 
the 
United States Department
of Energy.

\end{document}